\documentstyle[12pt]{article}

\def\beq{\begin{equation}}
\def\eeq{\end{equation}}
\def\barr{\begin{eqnarray}}
\def\earr{\end{eqnarray}}

\def\b{\bigskip}

\begin{document}

\title{Inhomogeneous Condensates in Planar QED\footnote{hep-th/9511192}}

\author{\normalsize{Gerald Dunne and Theodore Hall} \\
\normalsize{Physics Department}\\
\normalsize{University of Connecticut}\\
\normalsize{Storrs, CT 06269 USA}\\}

\date{}

\maketitle

\begin{abstract}
We study the formation of vacuum condensates in $2+1$ dimensional QED in the
presence of inhomogeneous background magnetic fields. For a large class of
magnetic fields, the condensate is shown to be proportional to the
inhomogeneous magnetic field, in the large flux limit. This may be viewed as a
{\it local} form of the {\it integrated} degeneracy-flux relation of Aharonov
and Casher.
\end{abstract}

\section{Introduction}

Parity-- and flavor--symmetry breaking aspects of $2+1$ dimensional QED have
been the subject of much research in recent years. This subject has
applications in planar electron systems and also provides a deep analogue of
certain features of symmetry breaking in $3+1$ dimensional theories relevant
for particle physics \cite{niemi,pisarski,appelquist}. An important focus of
these studies is the question of induced charges and spins, which are also
related to induced vacuum condensates. Gusynin et al \cite{miransky} have shown
recently that a uniform background magnetic field acts as a catalyst for
dynamical flavor symmetry breaking in $2+1$ dimensions. A key part of this
argument is the appearance of a nonzero vacuum flavor condensate, in the limit
of zero fermion mass, in the presence of a uniform background magnetic field of
strength $B$:
\beq
<0 |\bar{\psi}\psi |0>|_{m\to 0}=-sign(m){B\over 2\pi}
\label{c}
\eeq
While much can be learned from this constant $B$ field case, in order to
include dynamical gauge fields it is desirable to have a more complete
understanding of this phenomenon for more general background electromagnetic
fields. As (\ref{c}) refers to a uniform $B$ field, it is of course consistent
with the {\it integrated} relation
\beq
\int d^2x\, <0 |\bar{\psi}(\vec{x})\psi (\vec{x}) |0>|_{m\to 0}=
-sign(m){1\over 2\pi}\int d^2x\, B(\vec{x})=-sign(m)\Phi
\label{d-f}
\eeq
(where $\Phi$ is the net magnetic flux), which is essentially Landau's
degeneracy-flux relation \cite{landau}, and which was extended to inhomogeneous
magnetic fields by Aharonov and Casher \cite{aharonov,jackiw}. Much important
work has been done exploring the detailed global aspects of this integrated
result (\ref{d-f}), and relating it to mathematical index theorems
\cite{kiskis,stone,tan,boyanovsky,poly}.

The emphasis of this paper is rather different - here we investigate the extent
to which (\ref{d-f}) may be viewed as a {\it local} relation
\beq
 <0 |\bar{\psi}(\vec{x})\psi (\vec{x}) |0>|_{m\to 0}
\stackrel{\normalsize{?}}{=} -sign(m){1\over 2\pi}  B(\vec{x})
\label{c?}
\eeq
when the background magnetic field is inhomogeneous. This, and the closely
related issue of induced spin, have been addressed for the special case of an
Aharonov-Bohm flux string magnetic field \cite{jaroszewicz}. In this paper, we
consider the formation of a vacuum condensate in the presence of a more general
spatially inhomogeneous static background magnetic field. We present some
illustrative examples in which the condensate may be evaluated explicitly, and
then we show that for a large class of inhomogeneous magnetic fields the
condensate is proportional to the magnetic field [just as in (\ref{c?})], but
only in the large flux limit.

In Section II we give a brief review of vacuum condensates in $2+1$ dimensional
QED. Section III contains two explicit examples of particular inhomogeneous
magnetic fields and Section IV contains the general asymptotic analysis for
radial magnetic fields. Finally, we conclude with some brief comments.

\section{Vacuum Condensates in Planar QED}

Consider a parity invariant model of $2+1$ dimensional quantum electrodynamics
with fermionic Lagrange density
\beq
{\cal L}_F=\bar{\psi}\left(i\Gamma^\mu D_\mu -m\right)\psi
\label{lag}
\eeq
Here $\psi$ is a four-component spinor and the gamma matrices $\Gamma^\mu$
belong to a $4\times 4$ {\it reducible} representation
\beq
\Gamma^\mu=\left(\matrix{\gamma^\mu &0\cr0&-\gamma^\mu}\right) ,
\label{red}
\eeq
where the $2\times 2$ {\it irreducible} gamma matrices $\gamma^\mu$ are given
by
\barr
\gamma^0&=&\sigma^3=\left(\matrix{1&0\cr 0&-1}\right)\cr\cr
\gamma^1&=&i\sigma^1=\left(\matrix{0&i\cr i&0}\right)\cr\cr
\gamma^2&=&i\sigma^2=\left(\matrix{0&1\cr -1&0}\right)
\label{irred}
\earr
These gamma matrices are normalized as $\{\Gamma^\mu , \Gamma^\nu \}
=-2g^{\mu\nu} {\bf 1}$, where the flat Minkowski metric is
$g^{\mu\nu}=diag(-1,1,1)$. The covariant derivative operator is
$D_\mu=\partial_\mu-iA_\mu$, where for notational convenience we have absorbed
a factor of ``$e$'' into the gauge field $A_\mu$. This model is invariant under
the generalized parity transformation \cite{roman,pisarski,appelquist}
\beq
x^1\to-x^1, \hskip 1.25cm A_1(x^1, x^2)\to -A_1(-x^1, x^2), \hskip 1.25cm
\psi(x^1, x^2)\to\left(\matrix{0& \sigma^1\cr \sigma^1&0}\right)\psi (-x^1,
x^2)
\label{parity}
\eeq
and in the massless limit, $m\to 0$, has a global $U(2)$ flavor symmetry
corresponding to the interchange of the two $2\times 2$ irreducible
representations.

We consider static background gauge fields and work in the Weyl ($A_0=0$)
gauge. Then the Dirac equation, $(i\Gamma^\mu D_\mu -m)\psi=0$, block
diagonalizes as
\beq
\left(\matrix{E-m&-(D_1-iD_2)&0&0\cr(D_1+iD_2)& E+m&0&0\cr
0&0&E+m&-(D_1-iD_2)\cr 0&0&(D_1+iD_2)&E-m}\right)\psi=0
\label{dirac}
\eeq
which illustrates the fact that this reducible representation model is
equivalent to a theory describing two species of two-component spinors, one
with mass $+m$ and the other with mass $-m$. Without loss of generality, we
choose $m$ to be positive, and we also choose the net magnetic flux to be
positive.

The upper $2\times 2$ sub-block of the Dirac equation (\ref{dirac}),
corresponding to the positive mass species, is solved by a two-component spinor
\beq
\chi = e^{-iEt} \left(\matrix{f\cr -{(D_1+iD_2)\over E+m} f}\right)
\label{two-comp}
\eeq
when $E\neq -m$, and where $f(x,y)$ is a solution of the two-dimensional
partial differential equation
\beq
-(D_1-iD_2)(D_1+iD_2)f=\alpha^2 f
\label{schrodinger}
\eeq
with $\alpha^2=E^2-m^2$. Thus, there are solutions of positive and negative
energy, $E=\pm \sqrt{\alpha^2+m^2}$. When $|E|\neq m$, we can write these
solutions (including the appropriate normalization factors) as
\barr
\psi_{\{1\}}^{(\pm)}&=& e^{\mp i|E|t}\sqrt{{|E|\pm m\over
2|E|}}\left(\matrix{f\cr \mp{(D_1+iD_2)\over |E|\pm m} f\cr 0\cr
0}\right)\cr\cr\cr
\psi_{\{2\}}^{(\pm)}&=& e^{\mp i|E|t}\sqrt{{|E|\mp m\over
2|E|}}\left(\matrix{0\cr 0\cr f\cr \mp{(D_1+iD_2)\over |E|\mp m} f}\right)
\label{sols}
\earr
where $f$ satisfies the Schr\"odinger-like equation (\ref{schrodinger}).
Here, the subscript $\{1\}$ refers to species $\{1\}$ which corresponds to mass
$+m$, while species $\{2\}$ has mass $-m$. The superscripts $(\pm )$ refer to
the positive and negative energy solutions.

The threshold states, with $|E|=m$, are special and must be specified
separately. Indeed, already from (\ref{sols}) we see that for species $\{1\}$
we can have a positive energy solution with $|E|=m$, but the $1/\sqrt{|E|-m}$
factor excludes a negative energy threshold state of this form. By contrast,
for species $\{2\}$ we can have a negative energy solution of this form with
$|E|=-m$, while the $1/\sqrt{|E|-m}$ factor now excludes a positive energy
threshold state. This imbalance leads to an asymmetry in the spectrum of
states, and this asymmetry is the ultimate source of the interesting symmetry
breaking effects in planar QED.

The explicit threshold states are
\beq
\psi_{\{1\}}^{(0+)}=e^{-imt}\left(\matrix{f^{(0)}\cr 0\cr 0\cr 0}\right)\hskip
2cm
\psi_{\{2\}}^{(0-)}=e^{+imt}\left(\matrix{0\cr 0\cr f^{(0)}\cr 0}\right)
\label{threshold}
\eeq
where $f^{(0)}(x,y)$ satisfies the {\it first-order} threshold
equation\footnote{Note that we have excluded the potential threshold states of
the form $\psi_{\{1\}}^{(0-)}=e^{imt}\left(\matrix{0\cr g^{(0)}\cr 0\cr
0}\right)$ and $\psi_{\{2\}}^{(0+)}=e^{-imt}\left(\matrix{0\cr 0\cr 0\cr
g^{(0)}}\right)$, where $g^{(0)}$ satisfies $(D_1-iD_2)g^{(0)}=0$, because if
the solutions to (\ref{first}) are normalizable then these $g^{(0)}$ solutions
are not, and vice versa \cite{aharonov,jackiw}.}
\beq
(D_1+iD_2)f^{(0)}=0
\label{first}
\eeq
Now expand the fermion field operator in terms of creation and annihilation
operators for an orthonormal set of positive and negative energy modes from
(\ref{sols}) and (\ref{threshold}):
\beq
\Psi=\sum_{i=1}^2\sum\hskip -14pt \int_n \sum\hskip -14pt \int_p \left[
b_{n,p}\psi^{(+)}_{(i),n,p}+d^\dagger_{n,p} \psi^{(-)}_{(i),n,p}\right]
\label{op}
\eeq
where $b_{n,p}$ and $d_{n,p}$ are fermionic annihilation operators. The label
$n$ refers to the eigenvalue $\alpha^2_n$ of the equation (\ref{schrodinger})
and hence specifies the energy, while the label $p$ distinguishes between
degenerate states. Note that both $n$ and $p$ may take discrete and/or
continuous values, depending on the equations (\ref{schrodinger}) and
(\ref{first}) respectively. Also, note that the sum over the species in
(\ref{op}) is understood to include the positive energy threshold states for
species $\{1\}$ and the negative energy threshold states for species $\{2\}$.

The vacuum expectation value $<0|\bar{\Psi}\Psi |0>\equiv <0|\Psi^\dagger
\Gamma^0\Psi |0>$ is then given by
\beq
<0|\bar{\Psi}\Psi |0>=-\sum\hskip -14pt \int_p \, |f_p^{(0)}|^2 -2m\sum\hskip
-14pt \int_{n>0}\sum\hskip -14pt \int_p {|f_{n,p}|^2\over |E|}
\label{vacuum}
\eeq
where the first term on the RHS only involves threshold states, while the
second term involves all states with $|E|>m$. In the massless limit the second
term vanishes and the condensate is simply
\beq
<0|\bar{\Psi}(\vec{x})\Psi (\vec{x})|0>\left |_{m\to 0}\right.=-\sum\hskip
-14pt \int_p \, |f_p^{(0)}(\vec{x})|^2
\label{condensate-1}
\eeq
The minus sign on the RHS is due to the fact that the condensate is a vacuum
expectation value and so it is a sum over occupied negative energy states; and
with $m$ positive, the only negative energy threshold states correspond to
species $\{2\}$, for which the $\Gamma^0$ sub-block is $-\gamma^0$. Changing
the sign of $m$ corresponds to interchanging the two species, so one has
instead $+\gamma^0$. Thus, the condensate should be more precisely written as
\beq
<0|\bar{\Psi}(\vec{x})\Psi (\vec{x})|0>\left |_{m\to 0}\right.=-sign(m)
\sum\hskip -14pt \int_p \, |f_p^{(0)}(\vec{x})|^2
\label{condensate-2}
\eeq
Note that the condensate is determined entirely by the threshold states, which
solve the first-order equation (\ref{first}). We now consider several examples
in which this condensate may be computed explicitly.

We begin with the familiar case of a uniform background magnetic field
\cite{landau}. There is still gauge freedom of how we choose to represent the
corresponding vector potential. This choice of gauge will determine the precise
form of the threshold condition (\ref{first}) (as well as the eigenvalue
equation (\ref{schrodinger}) which determines the complete spectrum). In the
`linear gauge', with $\vec{A}=(0,Bx)$, the threshold state equation
(\ref{first}) has normalized solutions
\beq
f^{(0)}_p(x,y)=\left({B\over \pi}\right)^{1/4} e^{ipy}e^{-(p-B x)^2/(2B)}
\label{landau}
\eeq
where the degeneracy label $p$ takes continuous values corresponding to a plane
wave in the $y$ direction. Thus, it is trivial to evaluate the condensate to be
\beq
<0|\bar{\Psi}(\vec{x})\Psi (\vec{x})|0>\left |_{m\to
0}\right.=-sign(m)\sqrt{{B\over \pi}}\int {dp\over 2\pi} e^{-(p-B
x)^2/B}=-sign(m){B\over 2\pi}
\label{landau-c}
\eeq
In the `radial gauge', with $\vec{A}={B\over 2}(-y,x)$, the threshold state
equation (\ref{first}) has normalized solutions
\beq
f^{(0)}_p(x,y)=\sqrt{{1\over \pi p!}\left({B\over 2}\right)^{p+1}}z^p
e^{-B|z|^2/4}
\label{laughlin}
\eeq
where we have defined the complex variable $z=x+iy$, and the degeneracy label
$p$ now takes integer values $p=0,1,2,\dots$. Once again, it is trivial to
evaluate the condensate to be
\beq
<0|\bar{\Psi}(\vec{x})\Psi (\vec{x})|0>\left |_{m\to 0}\right.=-sign(m){B\over
2\pi}e^{-B|z|^2/2}\sum_{p=0}^\infty \left({B\over 2}\right)^p {|z|^{2p}\over
p!} =-sign(m){B\over 2\pi}
\label{laughlin-c}
\eeq
The answer is, of course, the same in each gauge. Also note that in each of
these cases, (\ref{landau-c}) and (\ref{laughlin-c}), the condensate is
independent of $\vec{x}$, as is expected for a uniform $B$ field. We now turn
to some less trivial cases in which the magnetic field is spatially {\it
inhomogeneous}.

\section{Inhomogeneous Magnetic Fields: Two Examples}

In this Section we consider two illustrative examples of specific inhomogeneous
 background magnetic fields. The first example is in the `radial gauge', for
which we choose the gauge field to be
\beq
\vec{A}=(-\partial_y \phi, \partial_x \phi )
\label{gauge}
\eeq
where $\phi=\phi(r)$ is some function only of the radial coordinate $r$. Then
the magnetic field is radial, $B(r)={1\over r}{d\over dr}(r{d\over dr} \phi)$,
and the (un-normalized but mutually orthogonal) solutions to the threshold
condition (\ref{first}) are
\beq
f^{(0)}_p=z^p e^{-\phi}
\label{put}
\eeq
where $p$ is a non-negative integer.

We now choose a particular functional form for $\phi$ which will permit the
explicit normalization of these states:
\beq
\phi(r)= {BR^2\over 4} log\left(1+ {r^2\over R^2}\right)
\label{radial}
\eeq
The corresponding radial magnetic field is
\beq
B(r)={B\over \left(1+{r^2\over R^2}\right)^2}
\label{radial-mag}
\eeq
which has a {\it finite} net flux
\beq
\Phi\equiv{1\over 2\pi}\int d^2x B(r)={BR^2\over 2}
\eeq
The constant $B$ represents the maximum value of the magnetic field, and $R$ is
a characteristic length scale associated with the spatial variation of the
magnetic field. In the infinite flux limit, with $R\to\infty$, this example
reduces to the constant $B$ field example.

With this radial choice (\ref{radial}) for $\phi$, the threshold states in
(\ref{put}) may be normalized, yielding
\beq
f^{(0)}_p(x,y)={1\over \sqrt{\pi}} \left({1\over R^2}\right)^{(p+1)/2} {1\over
\sqrt{\beta (p+1, \Phi-p-1)}} {z^p\over \left(1+{r^2\over R^2}\right)^{\Phi/2}}
\eeq
where $\beta(u,v)\equiv\Gamma(u)\Gamma(v)/\Gamma(u+v)$ is the beta function.
Having finite flux, this system displays the novel feature that only a finite
number of these states are localized and normalizable. Indeed, these states are
only localized for $p< [\Phi]$, where $[\Phi]$ is the largest integer less than
$\Phi$. The contribution to the condensate from these `bound' states
\footnote{The {\it integrated} condensate is proportional to the net magnetic
flux $\Phi$, with both localized and continuum states contributing
\cite{kiskis,stone,tan,boyanovsky,poly}. Here, for a local analysis of the
condensate {\it density} we only consider the localized bound states, in part
because the continuum states contribute at infinity, and also because in the
large flux limit the magnitude of their contribution is negligible.} is
\beq
<0|\bar{\Psi}(\vec{x})\Psi (\vec{x})|0>\left |_{m\to 0}\right.= -sign(m){1\over
\pi R^2} {1\over \left(1+{r^2\over R^2}\right)^\Phi} \sum_{p=0}^{[\Phi]}
{\left({r\over R}\right)^{2p}\over \beta (p+1, \Phi-p-1)}
\label{r-1}
\eeq
When $\Phi$ is an integer we can in fact evaluate this sum exactly, yielding
\beq
<0|\bar{\Psi}(\vec{x})\Psi (\vec{x})|0>\left |_{m\to 0}\right.=
-sign(m)\left(1-{1\over \Phi}\right){1\over 2\pi}{B\over \left(1+{r^2\over
R^2}\right)^2}
\label{radial-c}
\eeq
Note that the condensate has the same form as the magnetic field, but with an
overall multiplicative factor depending on the net magnetic flux $\Phi$. For
large $\Phi$ this factor approaches unity, so that
\beq
<0|\bar{\Psi}(\vec{x})\Psi (\vec{x})|0>\left |_{m\to 0}\right.\sim
-sign(m){B(r)\over 2\pi}
\label{rad-c}
\eeq

For our second example, consider a magnetic field that is uniform in one
direction (say the y-direction), and spatially varying in the $x$ direction. To
achieve this type of configuration, we choose a convenient `linear gauge' with
$\vec{A}=(0,a(x))$. The corresponding magnetic field, $B(x)=a^\prime (x)$, is
just a function of $x$, and the (un-normalized but mutually orthogonal)
solutions to the threshold equation (\ref{first}) are
\beq
f^{(0)}_p=e^{ipy}e^{-\int^x(a(x)-p)}
\label{lin-t}
\eeq
The specific choice $a(x)=\lambda B \, tanh(x/\lambda)$ yields a magnetic field
profile
\beq
B(x)={B\over cosh^2(x/\lambda)}
\eeq
The corresponding flux is
\beq
\Phi\equiv{1\over 2\pi}\int d^2x B={BL\lambda\over \pi}
\eeq
where we have compactified the $y$-direction with a length $L$. For this type
of magnetic field it is, in fact, possible to solve equation
(\ref{schrodinger}) for the {\it entire} spectrum, permitting for example the
{\it exact} evaluation of the effective energy \cite{cangemi}. Here, however,
for the evaluation of the vacuum condensate, we only need the normalized
threshold states
\beq
f_p^{(0)}(x,y)=\sqrt{{2\over L\lambda}}{1\over \sqrt{\beta(B\lambda^2+p\lambda
, B\lambda^2-p \lambda )}}\, e^{ipy}\, {e^{px}\over \left(2
cosh(x/\lambda)\right)^{B\lambda^2}}
\eeq
With the $y$-direction compactified, the degeneracy label $p$ takes discrete
values $p={2\pi k\over L}$, where $k$ is an integer such that
\beq
|k|< {BL\lambda\over 2\pi} \equiv{\Phi\over2}
\eeq
in order for these states to decay at infinity. We can therefore perform the
sum in (\ref{condensate-2}), yielding
\beq
<0|\bar{\Psi}(\vec{x})\Psi (\vec{x})|0>\left |_{m\to 0}\right.=-{2 sign(m)\over
L\lambda \left(2 cosh(x/\lambda )\right)^{2B\lambda^2}}
\sum_{k=-\left[{BL\lambda\over 2\pi}\right]}^{\left[{BL\lambda\over
2\pi}\right]} {e^{4\pi k (x/L)}\over \beta (B\lambda^2+2\pi k {\lambda\over L},
B\lambda^2-2\pi k {\lambda\over L})}
\label{linear-c}
\eeq
As $L\to\infty$ we can replace the sum over $k$ by an integral and find
\beq
<0|\bar{\Psi}(\vec{x})\Psi (\vec{x})|0>\left |_{m\to 0}\right.=-sign(m){B\over
\pi}{1\over \left(2 cosh(x/\lambda )\right)^{2B\lambda^2}} \int_{-1}^1 dt
{e^{2B\lambda^2 t(x/\lambda)}\over \beta (B\lambda^2(1+t), B\lambda^2(1-t))}
\label{integral}
\eeq
It is straightforward to plot this condensate for various values of the
dimensionless combination $B\lambda^2$. One finds that the condensate has the
same general `bell-shaped' form as $-sign(m)B(x)/2\pi$, but that the
correspondence is not exact. Nevertheless, for large $B\lambda^2$ (which
corresponds to large flux) we can use Stirling's formula to make an asymptotic
expansion of the inverse beta function in the integrand of (\ref{integral}) to
obtain
\barr
<0|\bar{\Psi}(\vec{x})\Psi (\vec{x})|0>\left |_{m\to 0}\right. &\sim&
-sign(m){B\over 2\pi}\sqrt{B\lambda^2\over \pi}
{1\over \left( cosh(x/\lambda)\right)^{2B\lambda^2}} \\\nonumber
&& \int_{-1}^1 dt \sqrt{1-t^2} exp\left[B\lambda^2\left(2
t\left({x\over\lambda}\right)-log(1-t^2)-t log\left({1+t\over
1-t}\right)\right)\right]
\label{asymp}
\earr
The remaining integral over $t$ is suited for an asymptotic expansion, for
large $B\lambda^2$, using Laplace's method \cite{carl}. For an integral of the
form
\beq
I(N)=\int ds \, \Psi(s) \,exp\left( N \Omega(s)\right)
\eeq
the large $N$ leading asymptotic behavior is dominated by a critical value
$s_c$ at which the exponent function $\Omega(s)$ has a maximum, and is given by
\beq
I(N)\sim \sqrt{2\pi\over -N\Omega^{\prime\prime}(s_c)}\, \Psi(s_c) \, exp
\left(N\Omega(s_c)\right)
\label{laplace}
\eeq
Applying Laplace's method to the $t$ integral in (\ref{asymp}), for which the
critical point is at $t_c=tanh(x/\lambda)$, we find the condensate to be
asymptotically proportional to the inhomogeneous magnetic field:
\beq
<0|\bar{\Psi}(\vec{x})\Psi (\vec{x})|0>\left |_{m\to 0}\right.\sim
-sign(m){B\over 2\pi}{1\over \left( cosh(x/\lambda
)\right)^2}=-sign(m){B(x)\over 2\pi}
\label{asymptotic}
\eeq

\section{Asymptotic Analysis for General $B(r)$}

At first sight, one might think that the asymptotic proportionality between the
condenstate and the inhomogeneous magnetic field found in (\ref{rad-c}) and
(\ref{asymptotic}) is due to the special form of the inhomogeneous magnetic
field chosen in these examples. For example, in each of these cases the
normalization factors may be computed exactly and are given by beta functions.
In general it is not possible to compute the normalization factors in closed
form. However, we show in this Section that this asymptotic analysis applies to
very general inhomogeneous magnetic fields $B(r)$ and $B(x)$. As the analysis
is very similar in the two cases, we concentrate on the radial case. In the
conclusion we make some comments concerning the general $B(x,y)$ case.

We choose the gauge field in the radial gauge (\ref{gauge}) with
$\phi=\phi(r)$. It is convenient to write
\beq
\phi(r)={BR^2\over 4} h\left({r^2\over R^2}\right)
\label{phi}
\eeq
where $R$ is some characteristic length scale, and $B$ together with the
dimensionless function $h$ are chosen so that the overall normalization gives
net flux $\Phi={BR^2\over 2}$. The special cases $h(\xi)=\xi$ and
$h(\xi)=log(1+\xi )$ have been considered in the preceeding Sections. Then, as
before, the threshold states are given by (\ref{put}). Note that these are
automatically mutually orthogonal for any $\phi(r)$ by virtue of the angular
integration. The vacuum condensate is
\beq
<0|\bar{\Psi}(\vec{x})\Psi (\vec{x})|0>\left |_{m\to 0}\right.=-sign(m)
\sum_{p=0}^{[\Phi]} {\left(r/R\right)^{2p}e^{-\Phi \, h(r^2/R^2)}\over N_p^2}
\label{ccc}
\eeq
where the normalization factors $N_p^2$ are given by
\beq
N_p^2=\pi R^2\int_0^\infty d\left({r^2\over R^2}\right)\,\left({r\over R}
\right)^{2p}e^{-\Phi \, h(r^2/R^2)}
\label{norm}
\eeq
For large flux, the sum over $p$ in (\ref{ccc}) may be replaced by an integral
from $0$ to $\Phi$, which may be re-expressed in terms of the rescaled variable
$t=p/\Phi$ as
\beq
<0|\bar{\Psi}(\vec{x})\Psi (\vec{x})|0>\left |_{m\to 0}\right.=-sign(m)\Phi
\int_0^1 dt\left({e^{\Phi(tlog\xi-h(\xi))}\over N_t^2}\right)
\label{cc}
\eeq
with normalization factors
\beq
N_t^2=\pi R^2\int_0^\infty d\xi \,e^{\Phi(tlog\xi-h(\xi))}
\eeq
where $\xi={r^2\over R^2}$.

To evaluate the asymptotic form of the condensate (\ref{cc}) for large flux
$\Phi$ we first need the asymptotic form of the normalization factors for large
$\Phi$. This can also be done by Laplace's method, with the dominant
contribution coming from the maximum of the exponent function
\beq
\chi(\xi)=t\, log\xi-h(\xi)
\eeq
This defines a critical point $\xi_c=\xi_c(t)$, as a function of the parameter
$t$, via the implicit relation
\beq
\xi h^\prime (\xi)=t
\label{imp}
\eeq
Using (\ref{laplace}) we see that
\beq
N_t^2\sim \pi R^2\sqrt{{2\pi \over \Phi}}e^{\Phi[t log\xi_c(t)-h(\xi_c(t))]}
{1\over \sqrt{{t\over \xi_c(t)^2}+h^{\prime\prime}(\xi_c(t))}}
\eeq
Inserting this into the integrand in (\ref{cc}) we obtain
\beq
<0|\bar{\Psi}(\vec{x})\Psi (\vec{x})|0>\left |_{m\to 0}\right.\sim -sign(m)
{B\over 2\pi} \sqrt{{\Phi\over 2\pi}} \int_0^1 dt \sqrt{{t\over
\xi_c(t)^2}+h^{\prime\prime}(\xi_c(t))}
e^{\Phi[tlog(\xi/\xi_c(t))-h(\xi)+h(\xi_c(t))]}
\label{cccc}
\eeq
The remaining $t$ integral may also be expanded asymptotically using Laplace's
method, with exponent function
\beq
\Omega(t)=tlog\xi-tlog\xi_c(t)-h(\xi)+h(\xi_c(t))
\eeq
This leads to a maximum at the critical point $t_c=t_c(\xi)$ as a function of
$\xi=r^2/R^2$, defined by the relation $\xi=\xi_c(t_c)$. Applying the inverse
function theorem (see comments below equation (\ref{ccccc})), we find
\beq
t_c=\xi h^\prime (\xi)
\eeq
and
\beq
-\Omega^{\prime\prime}(t_c)= {1\over \xi(h^\prime(\xi)+ \xi
h^{\prime\prime}(\xi))}
\eeq
Combining all these pieces, several remarkable cancellations occur, and the
leading asymptotic form of (\ref{cccc}) is simply given by
\barr
<0|\bar{\Psi}(\vec{x})\Psi (\vec{x})|0>\left |_{m\to 0}\right.&\sim& -sign(m)
{B\over 2\pi} \left(h^\prime(\xi)+ \xi h^{\prime\prime}(\xi)\right) \cr\cr
&=&-sign(m){B(r)\over 2\pi}
\label{ccccc}
\earr
This result depends on subtle cancellations between the asymptotic expansion of
the (inverse of the) normalization integral (which is an integral over $r$ or
equivalently $\xi$) and the asymptotic expansion of the sum over the threshold
states (which becomes an integral over $t$). These cancellations rely on the
use of the inverse function theorem which assumes that the function $\xi
h^\prime(\xi)$ is one-to-one (see (\ref{imp})). This means that its derivative,
$(\xi h^\prime(\xi))^\prime$, has a fixed sign, which moreover must be positive
in order for the critical point to be a maximum. Since $(\xi
h^\prime(\xi))^\prime$ is just the magnetic field, we see that our result
applies over a region of space in which the inhomogeneous magnetic field is
positive. This is the relevant physical set-up and is consistent with our
intuitive expectation - in a region with positive magnetic field, the spin
density tends to align with the magnetic field; moreover, if the flux in this
region is very large, the {\it local} spin density will actually be
approximately proportional to the {\it local} inhomogeneous magnetic field. The
proportionality factor is such that when this result is integrated over the
region, we regain the Aharonov-Casher relation \cite{aharonov,jackiw} between
the integrated spin and the net magnetic flux.

\section{Concluding Remarks}

In this paper we have studied the vacuum condensate of parity invariant $2+1$
dimensional QED in the presence of inhomogeneous magnetic fields. After
presenting two explicit illustrative examples in which the condensate may be
evaluated in detail, we showed that in the limit of large flux the condensate
is {\it locally} proportional to the inhomogeneous magnetic field:
\beq
<0|\bar{\Psi}(\vec{x})\Psi (\vec{x})|0>\left |_{m\to 0}\right.\sim
\cases{-sign(m) {B(r)\over 2\pi} \cr\cr -sign(m){B(x)\over 2\pi}}
\eeq
for general physically relevant magnetic fields which either depend only on the
radial coordinate or only on one of the Cartesian coordinates. These relations
represent a {\it local} analogue of the {\it integrated} Aharonov-Casher
relation. To generalize this result to general static magnetic fields $B(x,y)$,
we note that the background vector potential may be represented as
\beq
\vec{A}=(-\partial_y \phi(x,y), \partial_x \phi(x,y))
\eeq
where
\beq
\nabla^2 \phi(x,y)=B(x,y)
\eeq
Then the threshold solutions are simply
\beq
f_p(x,y) = F_p(z) e^{-\phi(x,y)}
\eeq
where $F_p(z)$ is some holomorphic function. However, to proceed one must
construct an {\it orthogonal basis} of such holomorphic functions. When
$\phi=\phi(r)$ this may be achieved by the choice $F_p(z)=z^p$, as in
(\ref{put}), and when $\phi=\phi(x)$ by the choice $F_p(z)=e^{pz}$, as in
(\ref{lin-t}). For a general $\phi=\phi(x,y)$ the natural choice of orthogonal
basis is not so clear (a Gramm-Schmidt orthogonalization is too clumsy for the
subsequent summation). Further clarification of this issue should lead to
interesting insights into the properties of vacuum condensates of $2+1$
dimensional QED with dynamical gauge fields.

\b\b

\noindent{\bf Acknowledgement:} This work has been supported by the DOE grant
DE-FG02-92ER40716.00.



\end{document}